\documentclass[aps,prl,twocolumn,showpacs,superscriptaddress,floatfix,nofootinbib]{revtex4-1}
\usepackage{amsfonts}
\usepackage{graphicx}
\usepackage{amsmath}
\usepackage{times}
\usepackage{amssymb}
\usepackage{color}
\usepackage[colorlinks,bookmarks=false,citecolor=blue,linkcolor=red,urlcolor=blue]{hyperref}
\usepackage{bm}

\begin{document}

\title{
Crystals of bound states in the magnetization plateaus of the Shastry-Sutherland model
}

\author{Philippe Corboz}
\affiliation{Theoretische Physik, ETH Z\"urich, CH-8093 Z\"urich, Switzerland}
\author{Fr\'ed\'eric Mila}
\affiliation{Institut de th\'eorie des ph\'enom\`enes physiques, \'Ecole Polytechnique F\'ed\'erale de Lausanne (EPFL), CH-1015 Lausanne, Switzerland}

\date{\today}

\begin{abstract}
Using infinite projected entangled-pair states (iPEPS), we show that the Shastry-Sutherland model in an external magnetic field
has low-magnetization plateaus which, in contrast to previous predictions, correspond to crystals
of bound states of triplets, and {\it not} to crystals of triplets. The first sizable plateaus appear at magnetization 1/8, 2/15 and 1/6, 
in agreement with experiments on the orthogonal-dimer antiferromagnet SrCu$_2$(BO$_3$)$_2$, and they can be naturally understood 
as regular patterns of bound states, including the intriguing 2/15 one. 
We also show that, even in a confined geometry, two triplets bind into a {\it localized} bound state with $S_z=2$. Finally, we discuss the role of competing domain-wall 
and supersolid phases as well as that of additional anisotropic interactions. 
 \end{abstract}


\maketitle
Predicting the phases of frustrated spin systems is one of the major challenges in theoretical condensed matter physics~\cite{Lacroix11}. A famous example is the Shastry-Sutherland model (SSM)\cite{Shastry81}, which is believed to accurately capture the physics of the orthogonal-dimer antiferromagnet SrCu$_2$(BO$_3$)$_2$.  A big effort has been invested in understanding the appearance of various magnetization plateaus observed in experiments~\cite{Kageyama99,Onizuka00,kageyama00,Kodama02,takigawa04,levy08,Sebastian08,isaev09,Jaime12,takigawa13,matsuda13}. Early on it was found that the SSM has almost localized triplet excitations~\cite{Miyahara99,kageyama00} which suggests that each plateau corresponds to a particular crystal of localized triplets. This viewpoint has been supported by many analytical and (approximate) numerical studies over the last 15 years~\cite{Miyahara99,momoi00a,Momoi00,fukumoto00b,fukumoto01,Miyahara03,miyahara03b,Dorier08,Abendschein08,takigawa10,Nemec12,Lou12}.

The SSM is given by the Hamiltonian 
\begin{equation}
H=J\sum_{\langle i,j \rangle}\bm S_{i}\cdot \bm S_{j}+J'\sum_{\langle \langle i,j \rangle\rangle}\bm
S_{i}\cdot \bm S_{j}-h\sum_{i}S_i^z
\end{equation}
where the $\langle i,j \rangle$ bonds with coupling strength $J$ build an array of orthogonal dimers while
the  bonds with coupling $J'$ denote inter-dimer couplings, and $h$ the strength of the external magnetic field. 

\begin{figure}
\begin{center}
\includegraphics[width=1\columnwidth]{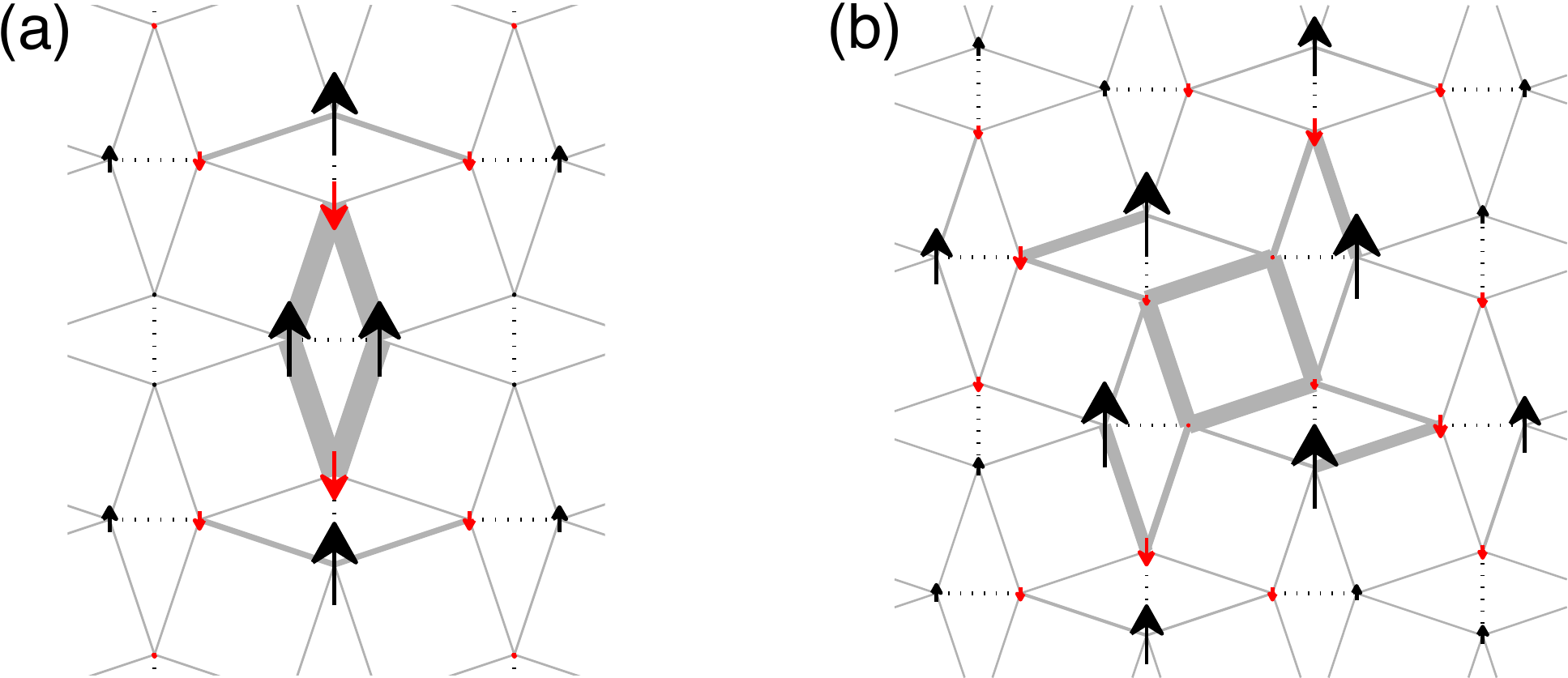}
\caption{(Color online) Spin structures of two types of excitations in the SSM obtained with iPEPS: (a) elementary triplet ($S_z=1$) excitation (obtained in a $6\times6$ unit cell), (b) bound state of triplets ($S_z=2$) forming a pinwheel pattern (obtained in a $4\times4$ unit cell). The thickness of the next-nearest neighbor bonds is proportional to the square of the bond energy (all the thick bonds have a negative energy).
}
\label{fig:states}
\end{center}
\end{figure}

The main result of this Letter is that - in contrast to the standard belief - the plateaus are not crystals of $S_z=1$ triplets (see Fig.~\ref{fig:states}(a)), but of $S_z=2$ bound states of triplets, which form a pinwheel pattern as shown in Fig.~\ref{fig:states}(b), and which are shown to be stable even if they are localized. Bound states have  previously been predicted to be relevant in the dilute limit of excitations~\cite{Momoi00, fukumoto00,knetter00,Totsuka01,knetter04}, but not for the formation of crystals. The only hint so far that bound states can form crystals was found in a one-dimensional analog - a SSM spin tube~\cite{manmana11b}.
It is also shown that the crystals formed by the bound states naturally explain the sequence of magnetization plateaus observed in SrCu$_2$(BO$_3$)$_2$. In particular, the  2/15 plateau is made of a simple and regular pattern of bound states, in contrast to the more complicated patterns of triplets which were previously suggested~\cite{Dorier08,Nemec12,takigawa13}. 


{\emph{Method  --} Our results have been obtained with infinite projected entangled-pair states (iPEPS) - a variational tensor-network ansatz to represent a two-dimensional wave function in the thermodynamic limit~\cite{verstraete2004,jordan2008,corboz2010}. It consists of a cell of tensors which is periodically repeated on the lattice, where in the present work we use one tensor per dimer. Each tensor  has 5 indices, a physical index for the local Hilbert space of a dimer, and four auxiliary indices which connect to the four nearest-neighboring tensors. Each tensor contains $4D^4$ variational parameters, where $D$ is the dimension of an auxiliary index called the bond-dimension which controls the accuracy of the ansatz. A $D=1$ iPEPS simply corresponds to a site-factorized wave function (a product state), and by increasing $D$ quantum fluctuations can be systematically added to the state. A $D=2$ iPEPS includes short-range quantum fluctuations and often qualitatively reproduces the results from linear spin-wave (or flavor-wave) theory~\cite{corboz11-su4,Corboz12_su4,Corboz13_su3hc}. Here we consider iPEPS with $D$ up to 12, which enables us to represent highly-entangled states.

By using different cell sizes iPEPS can represent different translational symmetry broken states (e.g. the different crystal structures). The tensors are either initialized randomly (this is how we found the bound states), or in a specific initial product state, e.g. corresponding to a particular triplet crystal. The latter states are typically metastable, i.e. we can increase $D$ and the state remains in the particular initial state. In this way we can compare the variational energies of different candidate ground states.

For more details on the method we refer to Refs.~\cite{Corboz13_shastry,matsuda13} where we used the same approach for the SSM at zero and at high magnetic fields. For the experts we note that the optimization of the tensors (i.e. finding the best variational parameters) has been done via an imaginary time evolution. Most of the results have been obtained with the so-called simple update~\cite{vidal2003-1,jiang2008},which gives a reasonably good estimate of the energy, but we checked several simulations with the more accurate (and computationally more expensive) full update (see Ref.~\onlinecite{corboz2010} for details). The contraction of the infinite tensor network is done with the corner-transfer matrix method~\cite{nishino1996,orus2009-1,Corboz13_shastry}. For the plateau phases we used tensors with U(1) symmetry to increase the efficiency~\cite{singh2010,bauer2011}.


{\emph{Ground state in the 1/8 plateau  --} We first focus on the 1/8 plateau in the physically relevant regime for SrCu$_2$(BO$_3$)$_2$, $J'/J=0.63$, as a first example to show that crystals made of bound states have a lower variational energy than crystals made of triplets.  
Two triplet crystals have been proposed: the diamond pattern shown in Fig.~\ref{fig:1o8}(a) with basis vectors $v_1=(2,-2)$ and $v_2=(2,2)$~\cite{miyahara03b,Nemec12,takigawa13}, and a rhomboid pattern defined by $v_1=(2,-2)$ and $v_2=(3,1)$ \cite{Kodama02,miyahara03b,takigawa04,isaev09} (see also Supplemental Material~\cite{SM}). In Fig.~\ref{fig:1o8}(c) we compare their variational energies with that of two different crystals of bound states: a square one with basis vectors  $v_1=(4,0)$, $v_2=(0,4)$~\cite{SM} and a rhomboid
one with basis vectors $v_1=(4,2)$, $v_2=(0,4)$ (Fig.~\ref{fig:1o8}(b)). For $D=2$ the diamond pattern of triplets has the lowest variational energy. This indicates that triplet crystals are favored if only low-order quantum fluctuations on top of a product state are taken into account. However, as soon as $D \geq 3$, the bound state crystals are energetically lower than the triplet crystals. The two bound state crystals are nearly degenerate, the  rhomboid one being slightly below the square one. 
Bound state crystals are also favored for other values of $J'/J$, as shown in the inset of Fig.~\ref{fig:1o8}(c).

\begin{figure}
\begin{center}
\includegraphics[width=1\columnwidth]{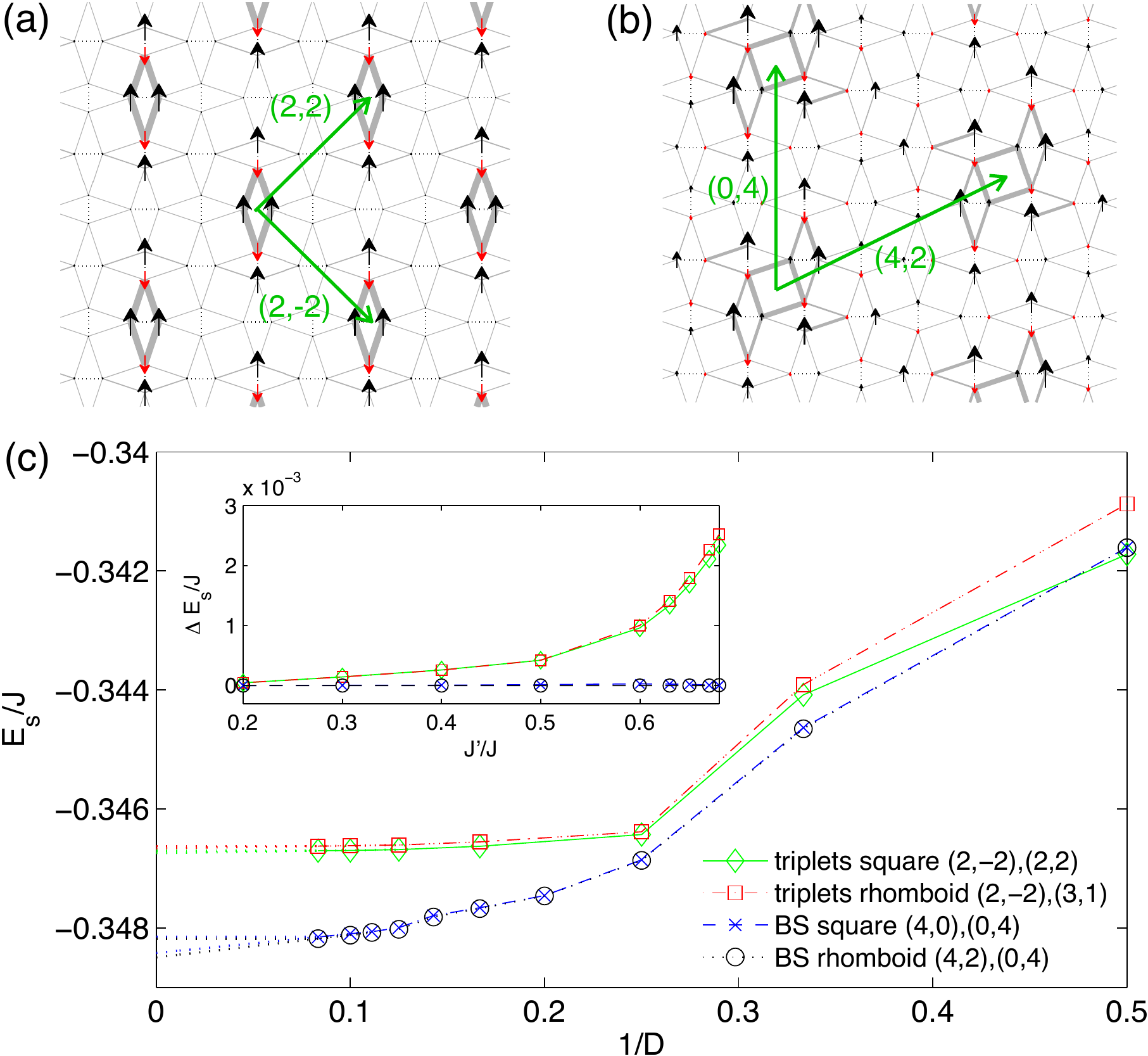}
\caption{(Color online) Spin structures of the candidate ground states for the 1/8 plateau phase: (a) the 1/8 square crystal made of triplets with magnetic unit cell vectors $v_1=(2,2)$, $v_2=(2,-2)$, (b) rhomboid crystal made of bound states (BS) defined by the vectors $v_1=(4,2)$ and $v_2=(0,4)$. 
(c)~Variational energies per site of the competing states as a function of the inverse bond dimension $1/D$ for $J'/J=0.63$ (the contribution from the Zeeman term has been subtracted). The unit cell vectors are given in brackets. The dotted lines are guides to the eye. The inset shows the energy difference with respect to the lowest energy state as a function of $J'/J$ for $D=8$.}
\label{fig:1o8}
\end{center}
\end{figure}


{\emph{Ground state in the 1/6 plateau  --} We made a similar study also for the 1/6 plateau~\cite{SM} and found that a bound state crystal with basis vectors $v_1=(6,0)$, $v_2=(0,4)$ (shown in Fig.~\ref{fig:magcurve}(d))  has clearly a lower energy than the previously proposed candidates of triplet crystals~\cite{Dorier08,Nemec12,takigawa13}. 



{\emph{Nature of the bound state and estimate of the binding energy  --}
The stabilization of crystals of $S_z=2$ bound states over $S_z=1$ triplet ones is in contradiction with the conclusions
of Ref.~\cite{Momoi00}, based on an expansion in $J'/J$, in which the authors argued that the $S_z=2$ bound state is only energetically favorable in the dilute limit thanks to the gain in kinetic energy via a correlated hopping process, and that in a crystalline phase where the bound state is localized, the two triplets would  actually repel each other. To make contact with Ref.~\cite{Momoi00}, we have calculated the binding energy of a localized bound state defined by $E_{bind}^{loc}=E_{bs}^{loc} - 2 E_{triplet}^{loc}$, where $E_{bs}^{loc}$ and $E_{triplet}^{loc}$ are the energies to form a single localized bound state and a localized triplet state, for $0\leq J'/J\leq 0.67$. These energies are estimated from simulations of a single bound state and triplet state in a $6\times6$ unit cell, where the repulsion between the triplets (or bound states) in neighboring cells is small, but the states remain localized within this cell~\cite{commentU1}. As can be seen in Fig.~\ref{fig:be}, this binding energy is negative for all ratios $J'/J$ as soon as $D$ is large enough, so that even in the perturbative regime of Ref.~\cite{Momoi00} we predict that there is a stable localized $S_z=2$ bound state. 

To resolve the apparent contradiction, we have revisited the calculation of Ref.~\cite{Momoi00}. More precisely, we have looked at the whole excitation spectrum in the two
triplet sector. It actually consists of four bands grouped into two pairs separated by a large gap~\cite{SM}. So, in the
spirit of Wannier functions for electronic bands, which are well localized if a band is separated from the others by a large gap, it must be
possible to reconstruct well localized wave functions from the lowest pair of bands, and the energy of these localized states will be 
of the order of the average energy of the bands. It turns out that the average energy is always {\it below} that of two triplets~\cite{SM}, leading to a binding energy of the same order as our estimate for small $J'/J$. So, we conclude that two triplets indeed bind into an $S_z=2$ bound state even in a confined geometry where the bound state is not allowed to delocalize.

That the bound state we discuss is related to the bound state discussed in Ref.~\cite{Momoi00} is further confirmed by 
its structure~\cite{SM}, which goes from two second neighbor 'dressed' triplets in the small $J'/J$ limit, as for the bound state of Ref.~\cite{Momoi00},
to the pinwheel structure of Fig.~\ref{fig:states}(b) for larger $J'/J$ with a central singlet plaquette reminiscent of the 
plaquette phase found at zero external magnetic field~\cite{Koga00,Takushima01,Chung01,Laeuchli02} for $0.675<J'/J<0.765$~\cite{Corboz13_shastry}. When $J'/J$ is large enough, there is actually an intuitive way 
to understand the stabilization of the
bound state: in a triplet excitation, a high energy cost has to be paid on the dimer with two parallel spins (cf. Fig.~\ref{fig:states}(b)). 
The bound state avoids this cost by distributing the four largest moments around a plaquette in such a way that they are neither nearest
nor next-nearest neighbors, while part of the energy lost in breaking four singlets on $J$ bonds is recovered by the formation of a singlet plaquette on $J'$ bonds.

\begin{figure}
\begin{center}
\includegraphics[width=1\columnwidth]{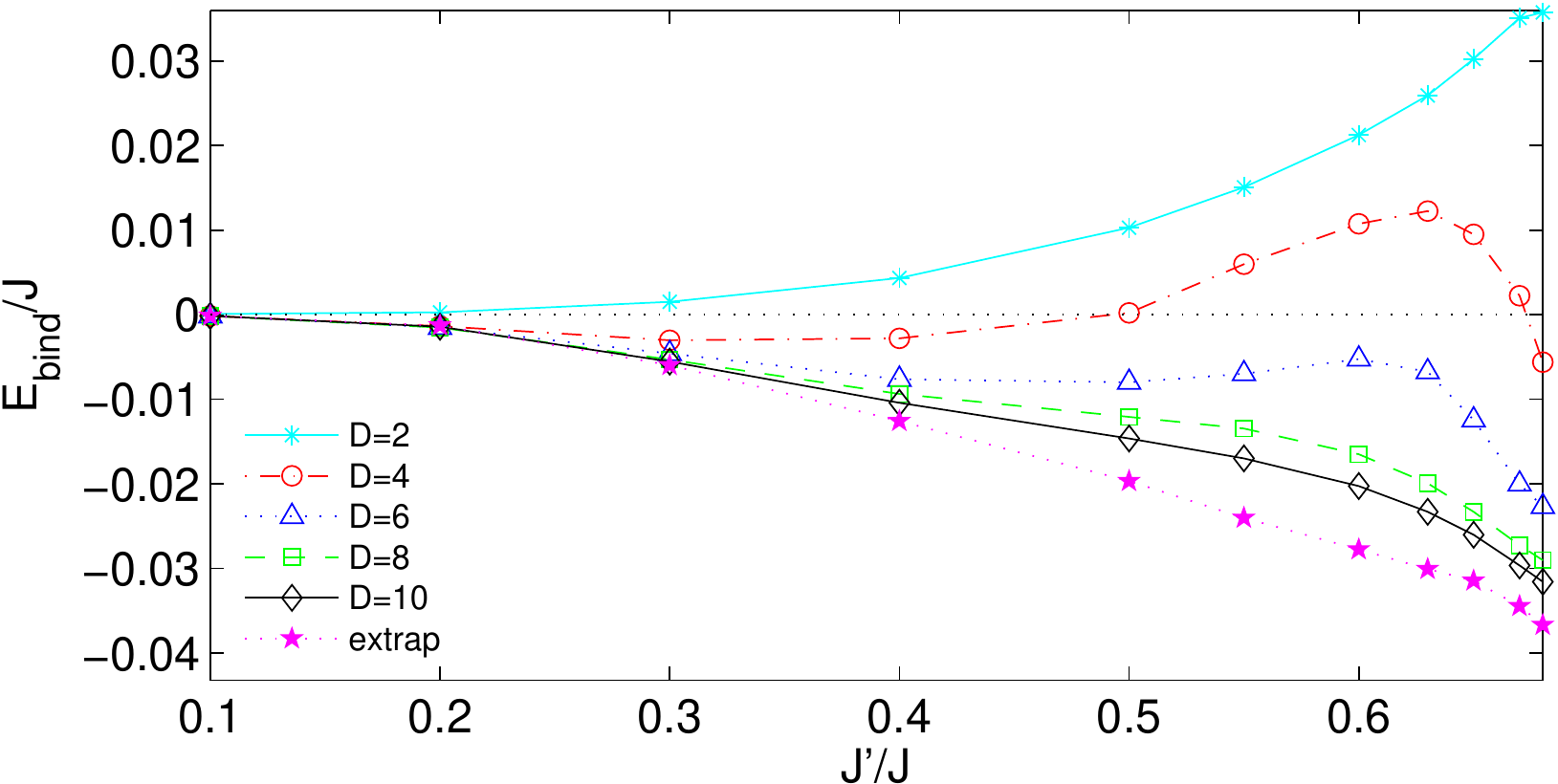}
\caption{(Color online) Estimate of the binding energy between two $S_z=1$ triplets as a function of $J'/J$, for different values of $D$.  }
\label{fig:be}
\end{center}
\end{figure}
{\emph{Magnetization curve  --}
Based on the previous findings it seems natural that different plateau states correspond to different crystals  of bound states. For each plateau with a magnetization $2/k$, $k$ integer, we have tested various structures (unit cell sizes) to determine the states with lowest energy. Then we have compared the variational energy of different plateaus, as a function of the external magnetic field $h$, to see which of the plateaus are energetically favored. The energy difference with respect to the lowest energy state as a function of $h$ for $D=10$ is plotted in Fig.~\ref{fig:magcurve}(a),
and the resulting magnetization curve in Fig.~\ref{fig:magcurve}(b). 
Sizable magnetization plateaus appear at  $1/8$, $2/15$, $1/6$, $1/5$, and $1/4$, besides the $1/3$ and $1/2$ plateaus at larger $h$~\cite{matsuda13}. The intermediate crystals 1/7, 2/13, 2/11, 2/9 are all higher in energy. The results are presented for $D=10$. The sequence is the same for other values of $D$, but the sizes of the individual plateaus are very sensitive to small changes in the energy 
and change with $D$~\cite{SM}.

\begin{figure}
\begin{center}
\includegraphics[width=1\columnwidth]{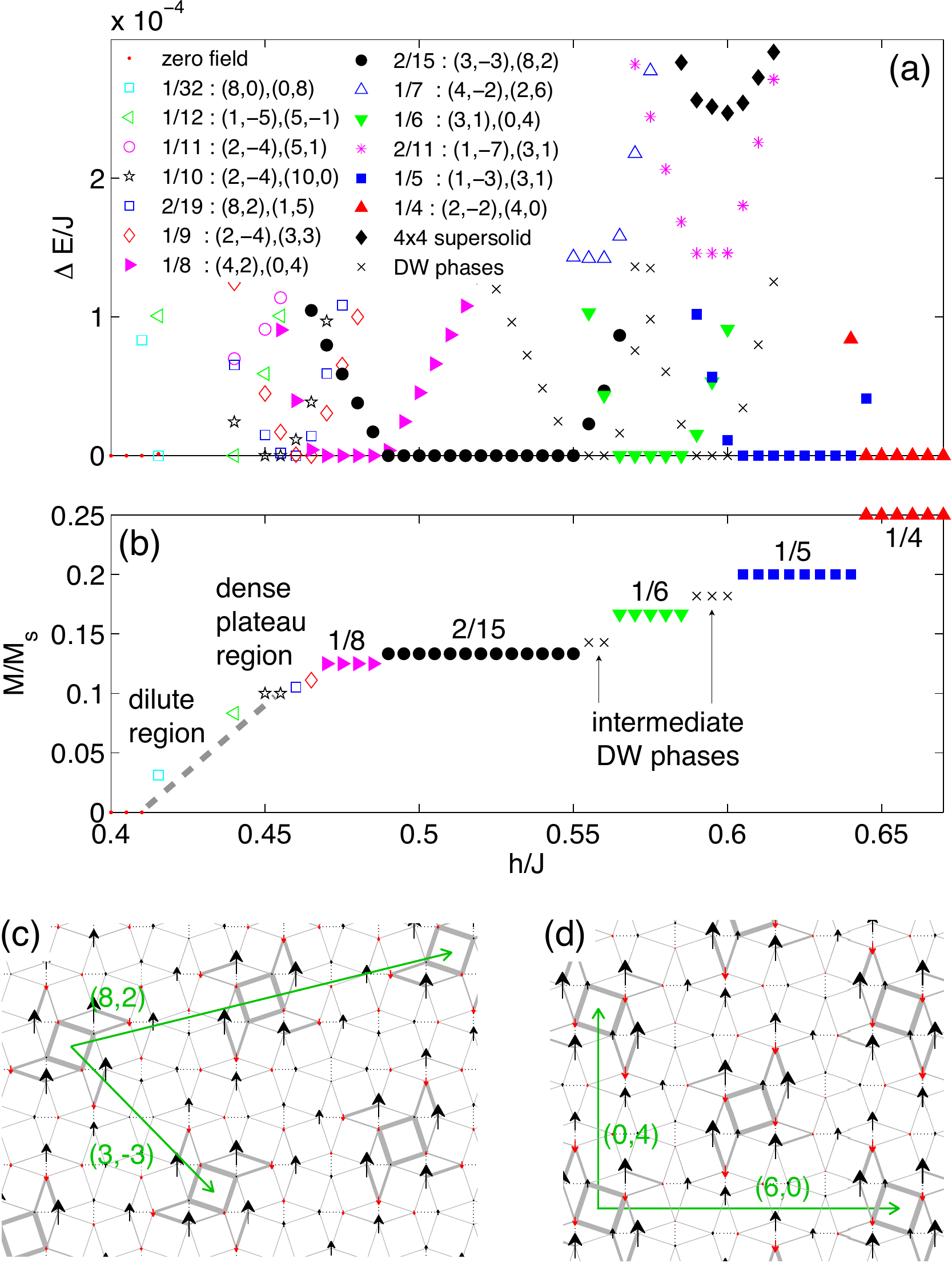}
\caption{(Color online) (a) Comparison of the variational energies of the various competing states as a function of $h$ with respect to the lowest energy state ($J'/J=0.63$ and $D=10$). The numbers in brackets in the legend show the unit vectors $v_1$, $v_2$ spanning the magnetic unit cell.  (b) Magnetization curve as a function of $h$ obtained from the lowest energy states.  Sizable plateaus are found for $1/8$, $2/15$, $1/6$, $1/5$, and $1/4$, besides smaller plateaus in the lower field region. (c)~Spin structure of the 2/15 plateau state. (d)~Spin structure of the 1/6 plateau state.
}
\label{fig:magcurve}
\end{center}
\end{figure}

Interestingly, the spin structure of the 1/4 plateau~\cite{SM} agrees with previous results. In fact, it can be seen either as a stripe of triplets \emph{or} as a stripe of bound states with unit cell vectors $v_1=(0,4)$, $v_2=(1,-1)$. The same is true for the 1/3 plateau (not shown).

Below the $1/8$ plateau we find a high density of plateaus which lie energetically very close, including plateaus at 1/12, 1/11, 1/10, 2/19, 1/9, see Fig.~\ref{fig:magcurve} and~\cite{SM}  (the 2/17 plateau lies slightly higher in energy). At even lower fields we enter the dilute region of bound states, where they start to delocalize (and eventually Bose condense). We did not study this region (marked by a dashed line in Fig.~\ref{fig:magcurve}(b)) in detail.

Besides the regular crystals of bound states we find domain-wall (DW) phases (crosses in Fig.~\ref{fig:magcurve}) between the 2/15 and the 1/6 plateau, and between the 1/6 and the 1/5 plateau. These DW phases are made of alternating domains (stripes) of the neighboring plateau states, with a magnetization depending on the individual widths of these domains. We restricted our study only to a few DW states, since these typically require very large cells. Other DW states for intermediate values of $M/M_s$ are likely to appear between the plateaus and further reduce the size of the adjacent plateaus. Eventually, a series of domain-wall states may connect the plateaus in a smooth way as observed experimentally~\cite{takigawa13}.

We have checked also for  competing supersolid phases in various cells, some of which have been found to be stable at higher magnetic fields~\cite{matsuda13}. For $J'/J=0.63$ the only close competing supersolid phase is given by the black diamonds in Fig.~\ref{fig:magcurve}(a), which is $>0.0002J$ higher than the plateau states. Its spin structure is shown in Fig.~\ref{fig:dm}(b).

Finally, we note that if we restrict our calculation to triplet crystals only, we reproduce the same sequence of magnetization plateaus as in Ref.~\cite{Dorier08} in the low-density limit of triplets ($M/M_s\le1/6$) (see \cite{SM}), but these states are always higher than the bound-state crystals.



\begin{figure}
\begin{center}
\includegraphics[width=1\columnwidth]{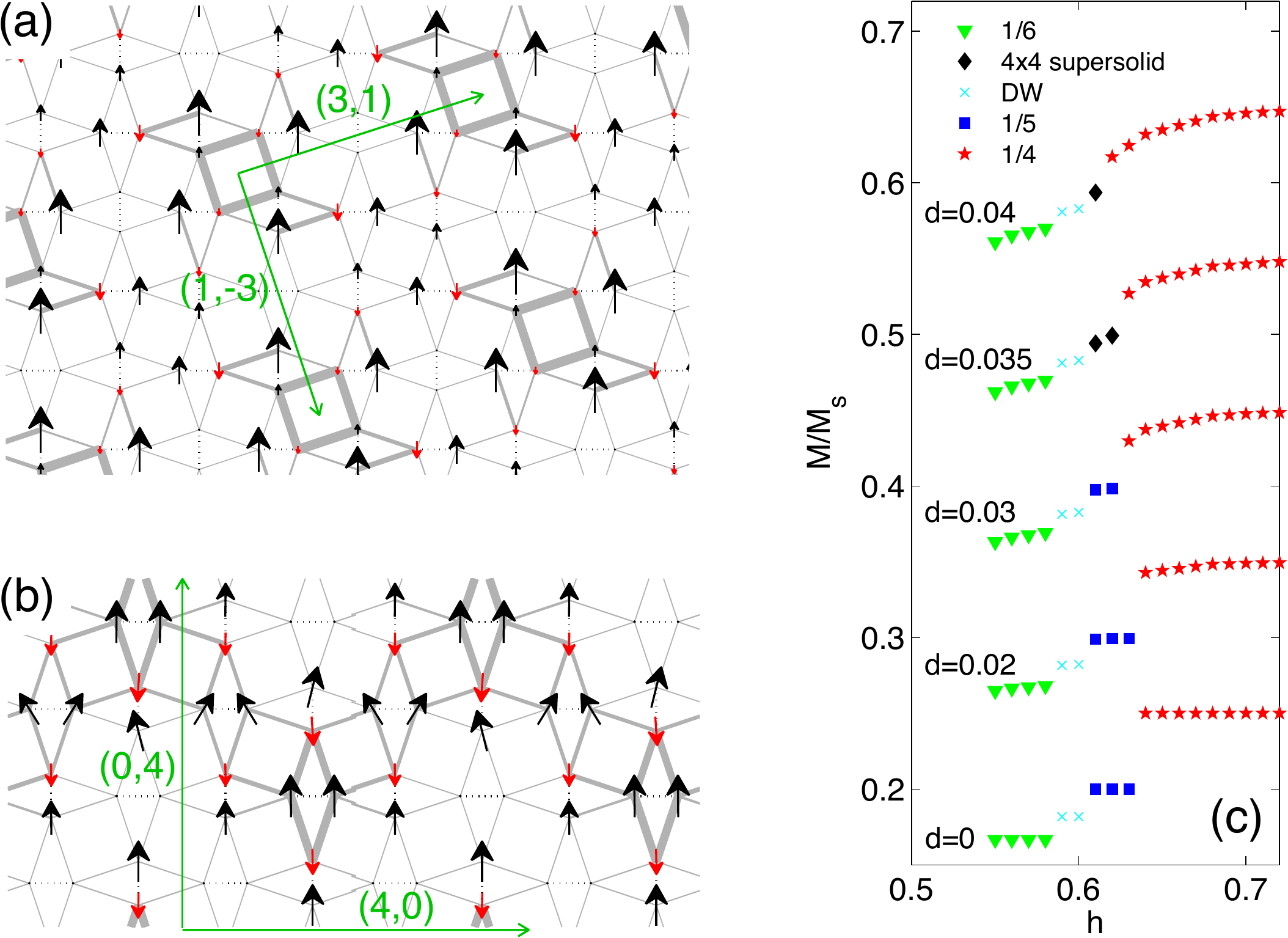}
\caption{(Color online) (a) Spin structure of the 1/5 plateau state. (b)~Spin structure of the supersolid phase in a 4x4 unit cell. (c)~Effect of a intra-dimer DM interaction with strength $d$ on the magnetization curve in the vicinity of the $1/5$ plateau ($D=6$): the plateaus start to acquire a finite slope, and the 1/5 plateau disappears due to the competing 4x4 supersolid phase shown in (b) and due to the competing (deformed) 1/4 plateau state. The individual magnetization curves are shifted by 0.1 for better visibility.
}
\label{fig:dm}
\end{center}
\end{figure}


{\emph{Discussion  --}
Our predicted sequence of magnetization plateaus is in good agreement with experiments on SrCu$_2$(BO$_3$)$_2$ up to 34T, where plateaus at 1/8, 2/15, 1/6, 1/4 have been found~\cite{takigawa13}. In particular, the 2/15 plateau appears very naturally as a regular crystal of bound states, as shown in Fig.~\ref{fig:magcurve}(c), in contrast to the more complicated patterns of triplets proposed previously. The size of the 2/15 plateau is large compared to the experimental data. However, it is likely to get further reduced by competing domain-wall phases in larger unit cells, which we did not consider in our simulations - based on our data we cannot give a precise estimate of the size of the individual plateaus.

At 1/5, there is no clear sign of a sizable plateau in experiments (only a discontinuity in the slope of the magnetization curve, see Ref.~\cite{takigawa13}), whereas our simulations predict a stable plateau, with the spin structure shown in Fig.~\ref{fig:dm}(a) (see \cite{SM} for a comparison with the structure found in Ref.~\cite{isaev09}). However, in the vicinity of the 1/5 plateau there is a competing supersolid phase which is only slightly higher in energy (Fig.~\ref{fig:dm}(b)). We have studied the effect of an additional intra-dimer 
Dzyaloshinskii-Moriya (DM) interaction $d$ on this competition and got the encouraging result that already a small but realistic value between $d=0.03J$ and $d=0.04J$ (cf. Refs.~\onlinecite{zorko04,room04,kodama05,takigawa10}) is enough to destabilize the 1/5 plateau in favor of the supersolid phase, see Fig.~\ref{fig:dm}(c). The systematic analysis including all anisotropic interactions  (intra- and inter-dimer DM interactions and g-tensor anisotropy \cite{Cepas01,kodama05,romhanyi11}) is left for future investigation.

In summary, our results are in strong support of an alternative explanation for the magnetization process in SrCu$_2$(BO$_3$)$_2$. In particular, the observed magnetization plateaus correspond to crystals of bound states and \emph{not} to crystals of triplets. Structures
are also proposed for the intermediate phases. The next
step will be to compare the resulting structures with NMR spectra~\cite{takigawa13}. We just note that the basic requirements (spins strongly polarized along the field and spins polarized opposite to the field surrounded by weakly polarized spins) are present in all the proposed structures.
%
Finally, our study also further demonstrates the potential of iPEPS as a powerful tool for open problems in frustrated magnetism. Thanks to (largely) unbiased simulations unexpected physics can be discovered.  The crystals of bound states found in this work definitely came as a surprise to us.

\begin{acknowledgments}
We would like to thank M. Takigawa for his critical reading of the manuscript. This work has been supported by the Swiss National Science Foundation. The  simulations have been performed on the Brutus cluster at ETH Zurich.
\end{acknowledgments}

\bibliographystyle{apsrev4-1}
\bibliography{refs,comments}

\begin{thebibliography}{58}
\bibitem [{com()}]{commentU1}%
  \BibitemOpen
  \href@noop {} {}\bibinfo {note} {In these simulations we enforce a U(1)
  symmetry on the tensors, which prevents the triplets or bound states from
  undergoing a Bose-Einstein condensation (i.e. from completely delocalizing).
  We note that new techniques to study localized excitations with tensor
  networks have been developed~\cite{Phien12,milsted13}, but so far they have
  only been tested in one dimension with matrix product states.}\BibitemShut
  {Stop}%
\end{thebibliography}

\end{document}


\title{
Crystals of bound states in the magnetization plateaus of the Shastry-Sutherland model: \textit{supplemental material}
}

\author{Philippe Corboz}
\affiliation{Theoretische Physik, ETH Z\"urich, CH-8093 Z\"urich, Switzerland}
\author{Fr\'ed\'eric Mila}
\affiliation{Institut de th\'eorie des ph\'enom\`enes physiques, \'Ecole Polytechnique F\'ed\'erale de Lausanne (EPFL), CH-1015 Lausanne, Switzerland}

\date{\today}
\maketitle

\section{Introduction}

In this supplemental material, we provide some additional information, mainly figures, about various aspects of the physics 
discussed in the main text. In Section II, we show sketches of the main bound state structures (stable crystals in II-A, higher energy
crystals in II-B, crystals of the dense plateau region of the low-density limit in II-C, and domain-wall structures in II-D). In section
III, we discuss the dependence of the magnetization curve
on the bond-dimension $D$. In Section IV, we provide a detailed comparison of the  bound state and triplet crystals at 
magnetization 1/6. 
Section V is devoted to the localized bound state out of which all the plateaus emerge (binding energy in V-A, local structure
in V-B). In Section VI we discuss the discrepancy between our results and those of Ref.~\onlinecite{isaev09} which have been obtained by a hierarchical mean-field approach. 
Finally, Section VII summarizes the information obtained with iPEPS on the triplet crystals (structures in VII-A and relative energy
in VII-B).

\section{Spin structures of the bound state patterns}

\subsection{Spin structure of the main stable bound state crystals}

\begin{figure}[h]
\begin{center}
\includegraphics[width=1\columnwidth]{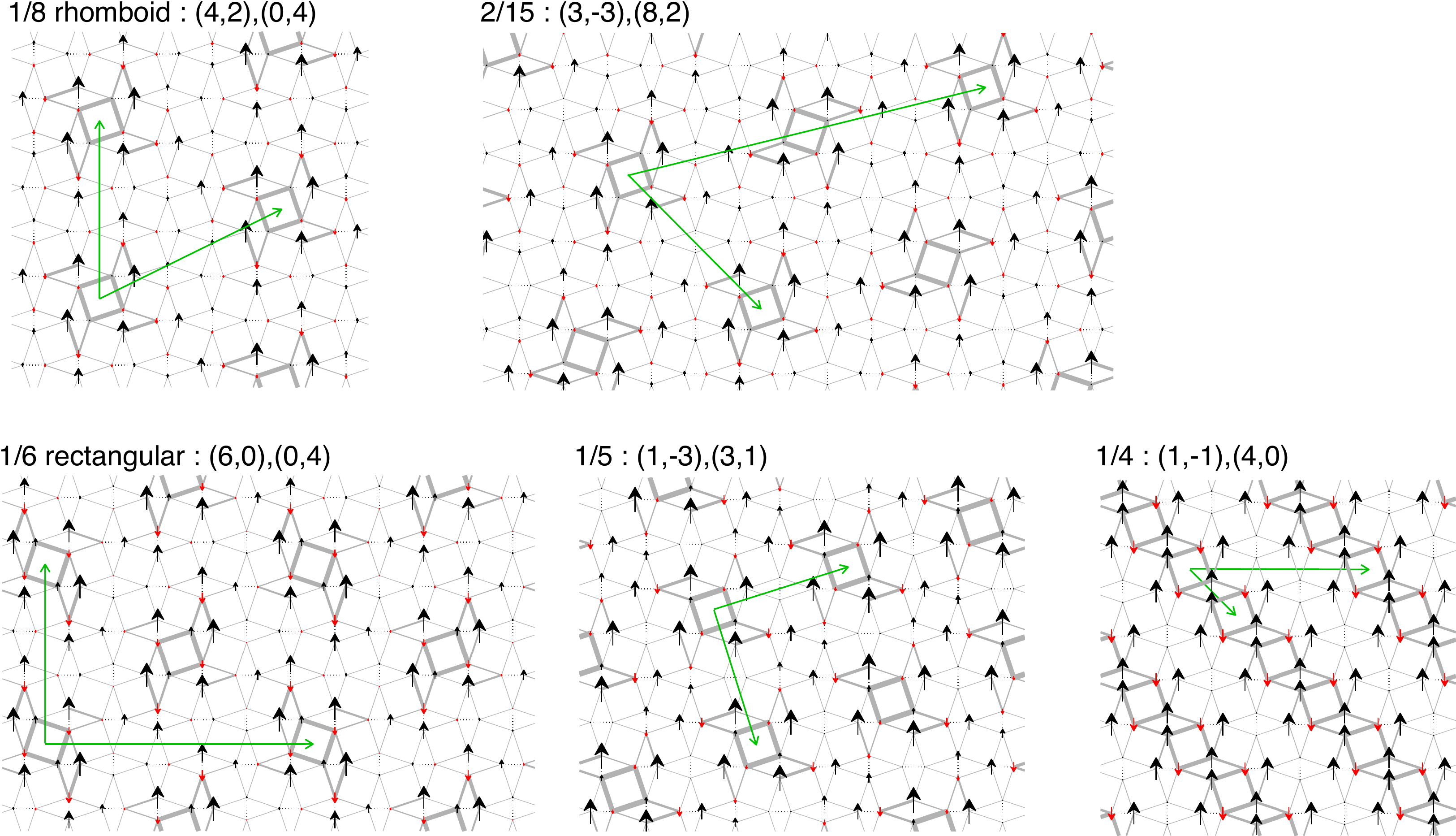}
\caption{Spin structures of the of the main stable bound state crystals at 1/8, 2/15, 1/6, 1/5 and 1/4 magnetization for $J'/J=0.63$ and $D=6$ (full update).}
\label{fig:mainplateaus}
\end{center}
\end{figure}

\clearpage
\subsection{Spin structure of higher-energy bound state crystals}

\begin{figure}[h]
\begin{center}
\includegraphics[width=1\columnwidth]{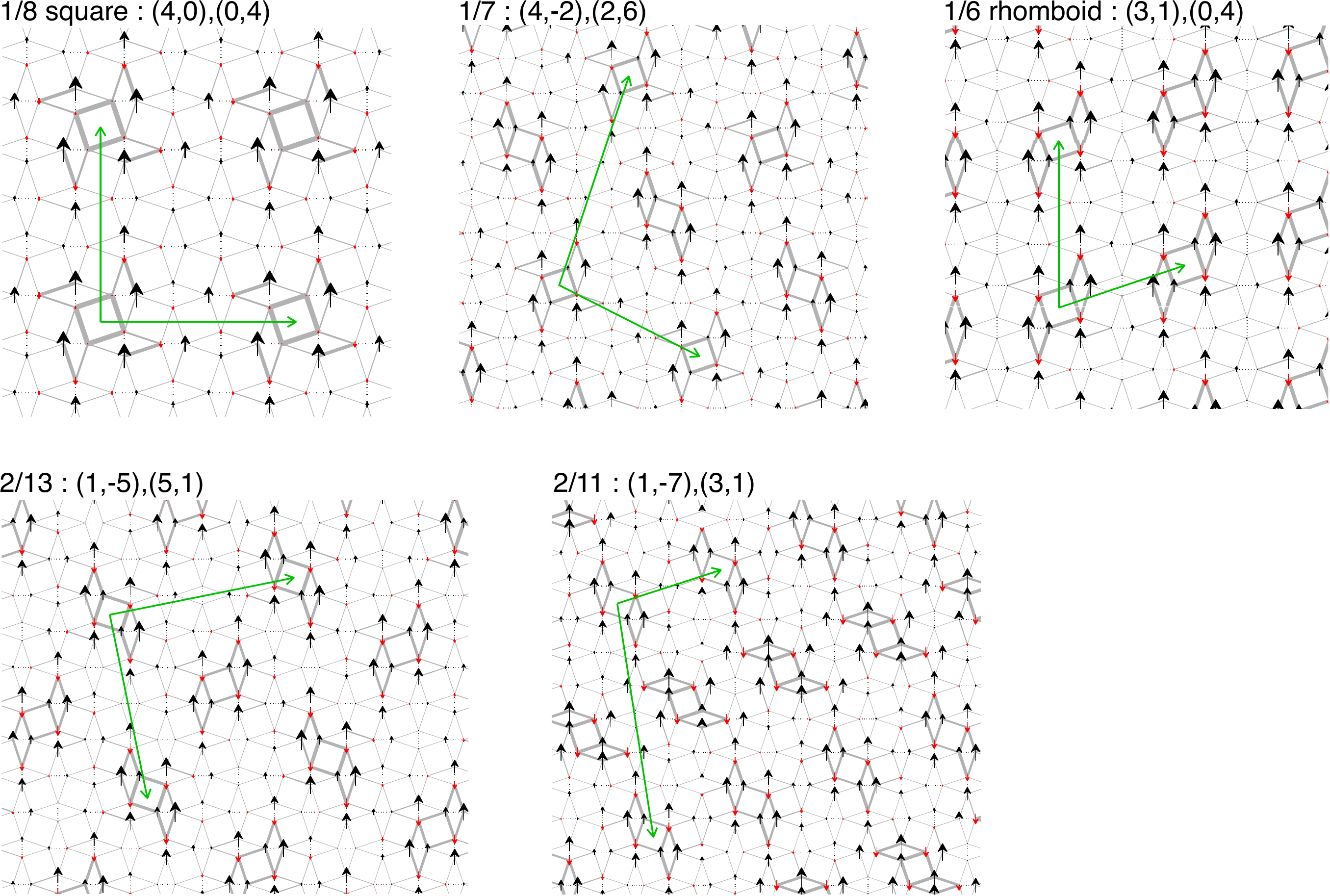}
\caption{Spin structures of higher-energy bound state crystals mentioned in the main text for $J'/J=0.63$ (simple update and $D=10$ , except for the 1/8 state where the full update was used). Note that the spin structures of the bound states obtained in simulations using the simple update appear less symmetric than the ones obtained with the full update.}
\label{fig:otherstates}
\end{center}
\end{figure}

\clearpage
\subsection{Spin structure of the bound state crystals in the dense plateau region (low magnetization)}

\begin{figure}[h]
\begin{center}
\includegraphics[width=1\columnwidth]{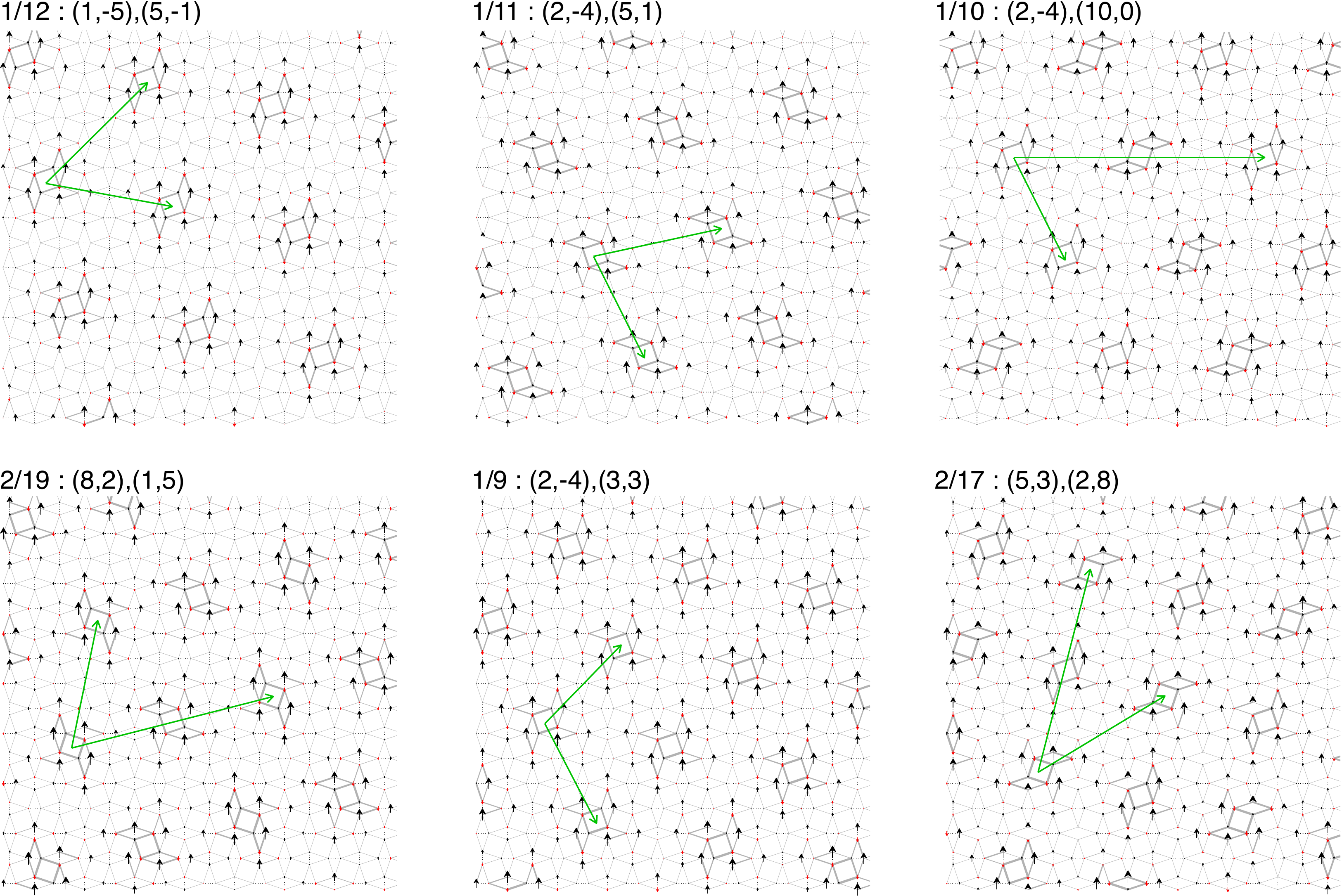}
\caption{Examples of spin structures of the of the bound state crystals below 1/8 for $J'/J=0.63$, $D=10$ (simple update).  }
\label{fig:denseplateaus}
\end{center}
\end{figure}

\clearpage

\subsection{Spin structure of the domain wall states}

\begin{figure}[h]
\begin{center}
\includegraphics[width=0.8\columnwidth]{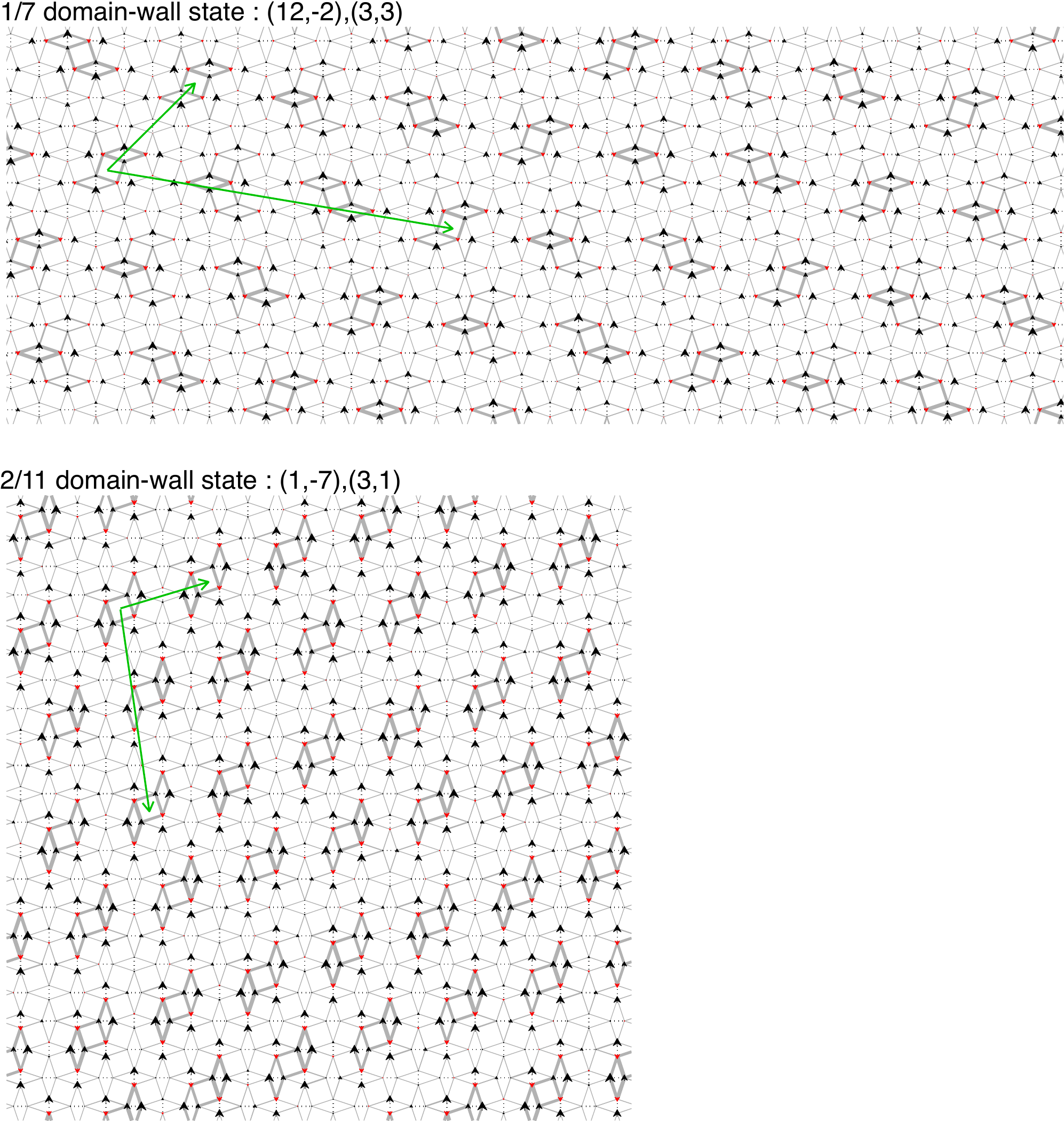}
\caption{Examples of spin structures of domain-wall phases, for $J'/J=0.63$ and $D=10$ (simple update). }
\label{fig:dwphases}
\end{center}
\end{figure}

\section{Magnetization curves for different values of $D$}
\begin{figure}[htb]
\begin{center}
\includegraphics[width=0.65\columnwidth]{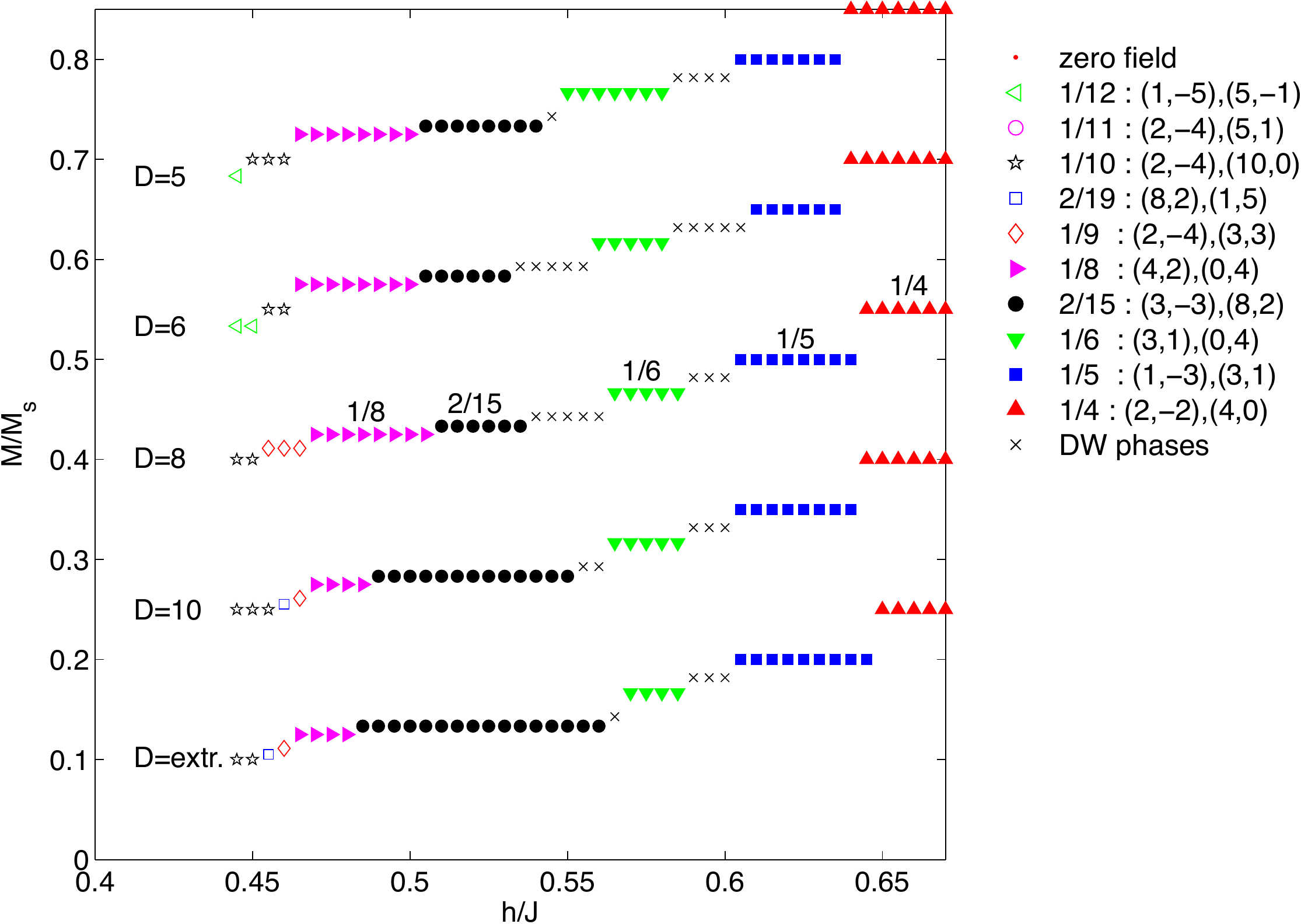}
\caption{Magnetization curves obtained for various values of $D$ and from an extrapolation in $1/D$ (simple update).}
\label{fig:manymagcurves}
\end{center}
\end{figure}

In Fig.~\ref{fig:manymagcurves} we show a comparison of the magnetization curves for different values of $D$. The sizes of the plateaus vary with $D$, but qualitatively the results are similar, in particular the same sequence of the sizable plateaus is found in all cases. As mentioned in the main text the sizes of the plateaus might be further decreased by other competing domain-wall phases.

\clearpage

\section{Bound state versus triplet crystals at magnetization 1/6}
\begin{figure}[htb]
\begin{center}
\includegraphics[width=0.6\columnwidth]{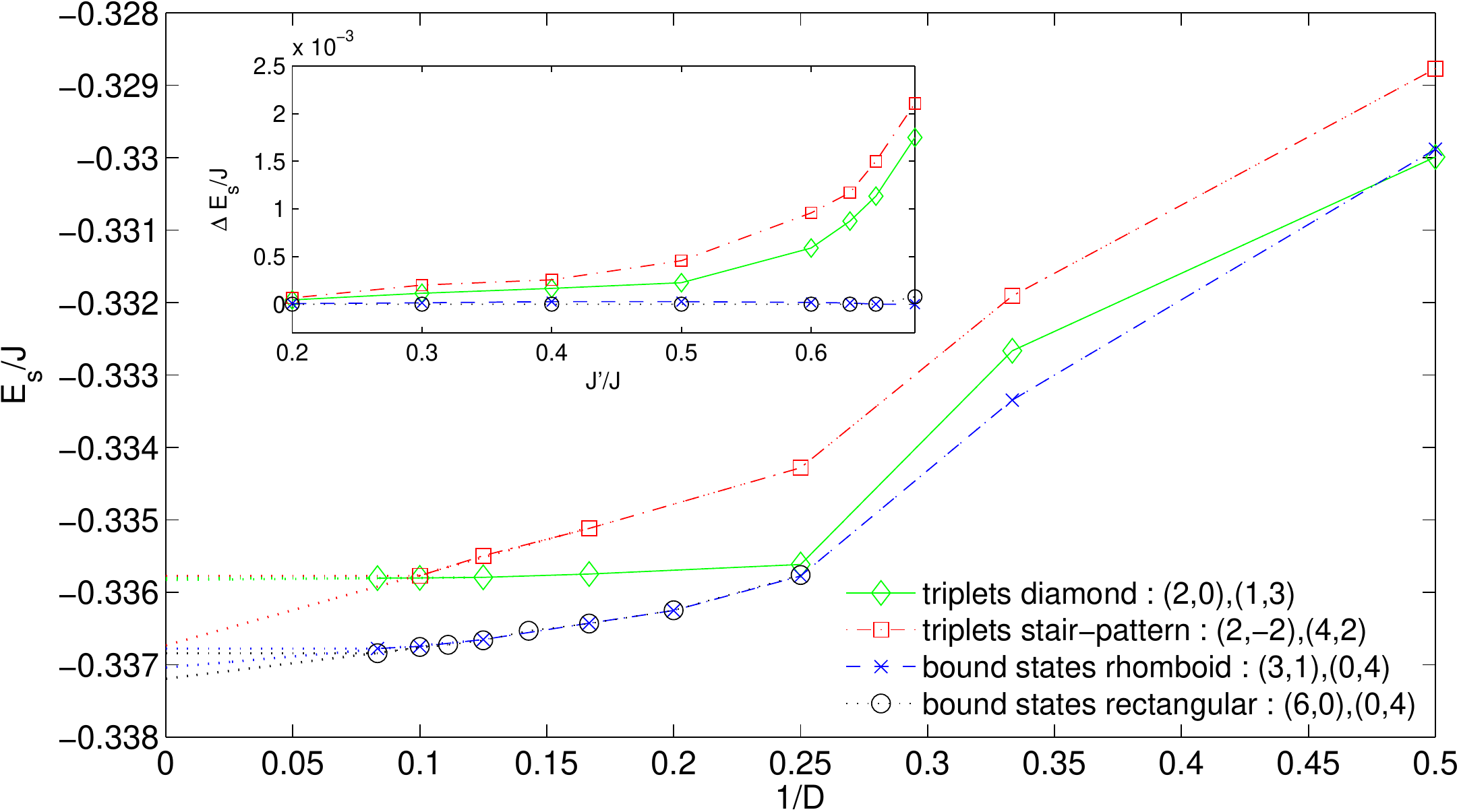}
\caption{Variational energies per site of the competing states of the 1/6 plateau as a function of inverse $D$ for $J'/J=0.63$ (the contribution from the Zeeman term has been subtracted). The inset shows the energy difference with respect to the lowest energy state as a function of~$J'/J$.}
\label{fig:1o6}
\end{center}
\end{figure}

In Fig.~\ref{fig:1o6} we show the variational energies of the competing states for the 1/6 plateau. We compare the previously proposed triplet candidates from Refs.~\onlinecite{Dorier08,Nemec12,takigawa13} (a diamond pattern defined by $v_1=(2,0)$, $v_2=(1,3)$ and  a stair-pattern given by $v_1=(2,-2)$, $v_2=(4,2)$, cf. Fig.~\ref{fig:tripletcrystals}) with two different bound state crystals (rhomboid with cell vectors $v_1=(3,1)$, $v_2=(0,4)$ (Fig.~\ref{fig:otherstates}) and rectangular with cell vectors $v_1=(6,0)$, $v_2=(0,4)$ (Fig.~\ref{fig:mainplateaus})) for $J'/J=0.63$. 
The two bound-state crystals have clearly a lower energy than the two triplet crystals, with the rectangular one slightly below the rhomboid one. %
A similar result is found also for other values of $J'/J$, as shown in the inset of Fig.~\ref{fig:1o6}.

The stair-pattern triplet crystal is still considerably decreasing with increasing $D$. However, we note that its structure is actually unstable, i.e. quantum fluctuations at large $D$ destroy the pattern, as shown in the last panel of Fig.~\ref{fig:tripletcrystals}.
So the fact that its energy still decreases very fast below $1/D=0.25$ does not mean that this triplet structure could ultimately be
stabilized, but that the variational ground state reached starting from this configuration is another configuration, presumably a 
bound state crystal.

\section{Localized bound state}

\subsection{Binding energy}

\begin{figure}[htb]
\begin{center}
\includegraphics[height=0.4\columnwidth]{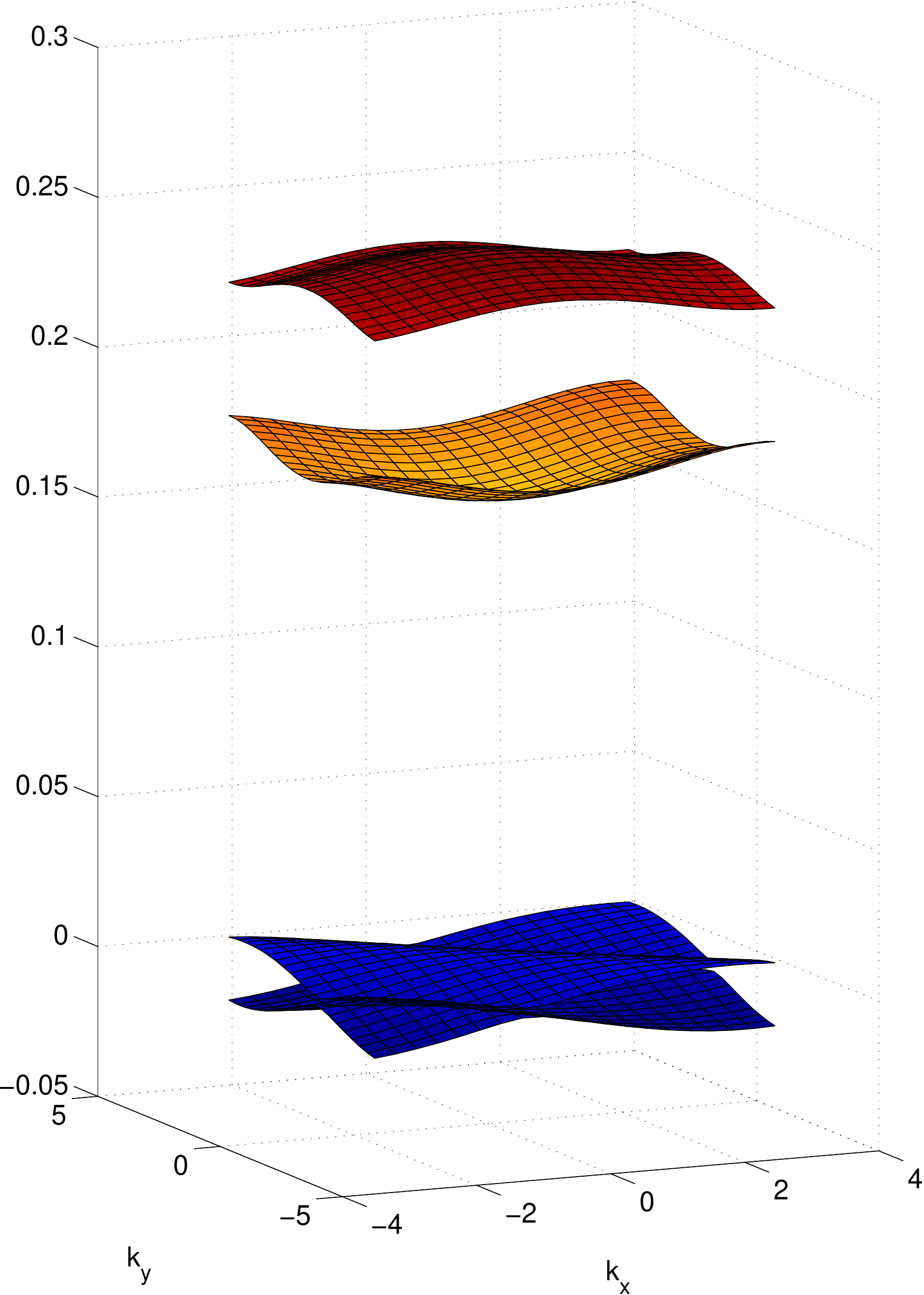} \hspace{0.5cm} 
\includegraphics[height=0.4\columnwidth]{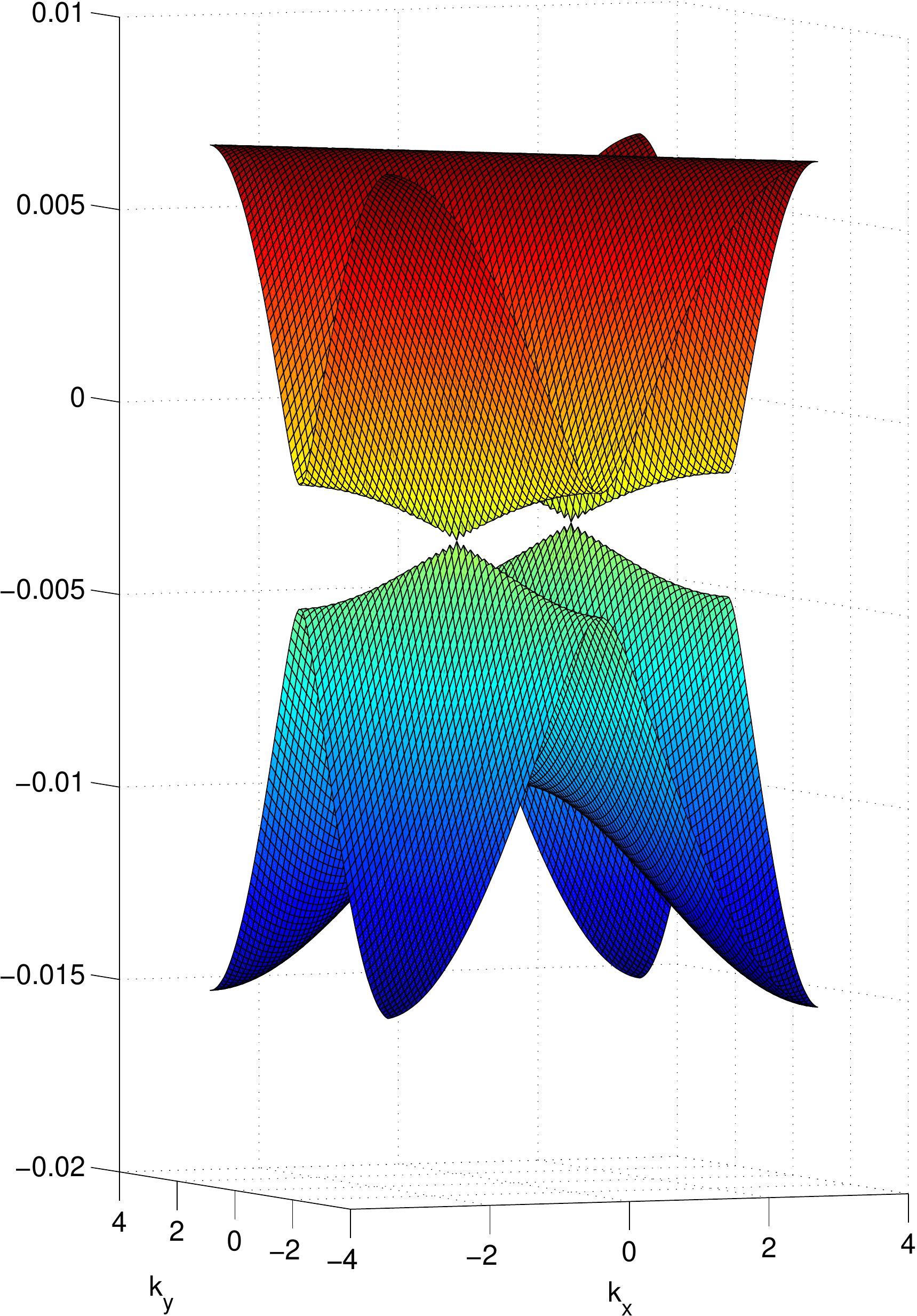} \hspace{0.5cm}
\includegraphics[height=0.4\columnwidth]{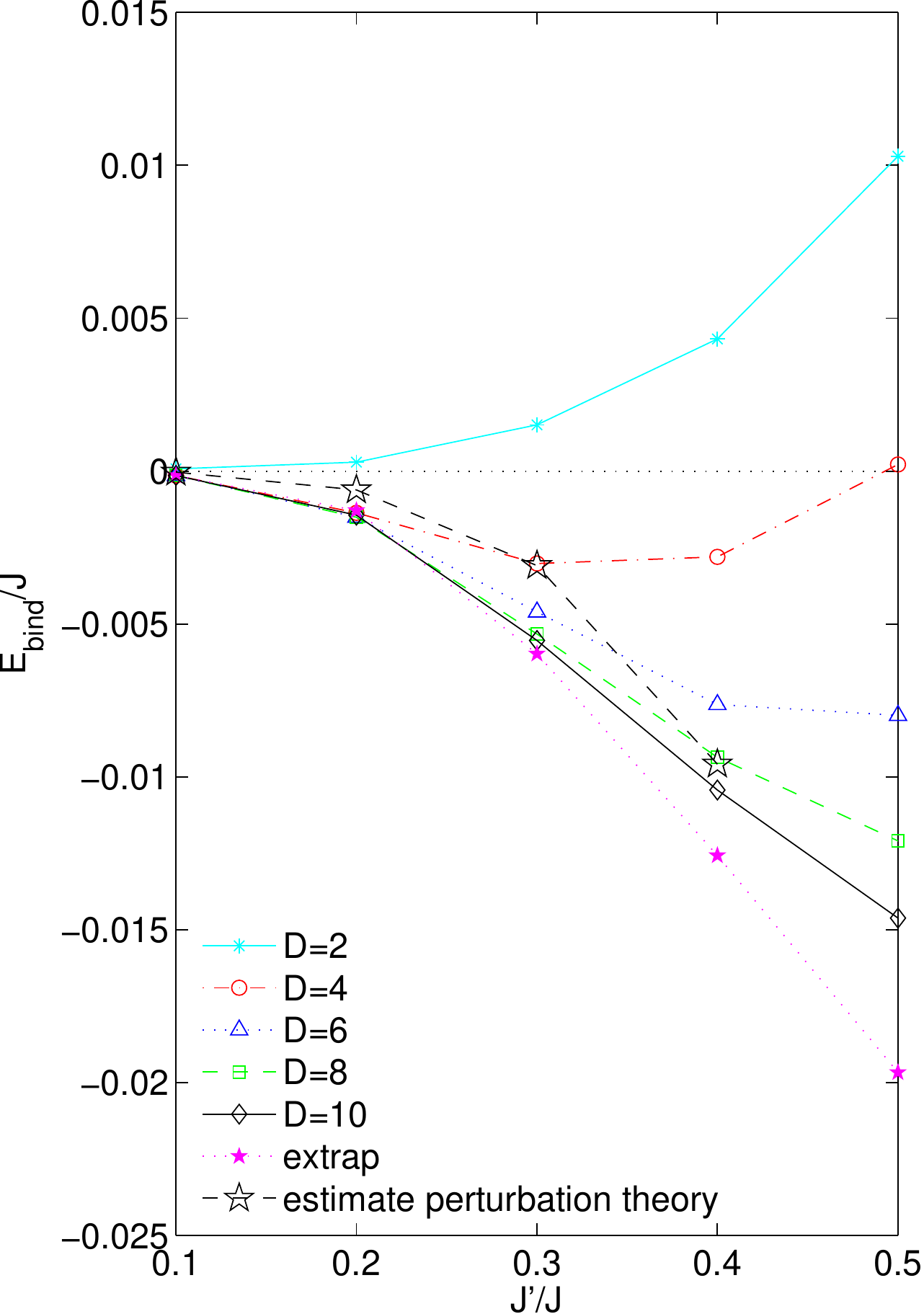}
\caption{Left panel: Band structure of the 4 bands for $J'/J=0.3$. The reference energy is that of two isolated triplets. The upper two bands are well separated from the lower two bands. Middle panel: Lower two bands for $J'/J=0.3$, which touch at two Dirac points. Right panel: Comparison of the iPEPS data for the binding energy with the estimate given by the mean-energy of the two lower bands as a function of $J'/J$.}
\label{fig:bandstructure}
\end{center}
\end{figure}

In this paragraph, we discuss the connection between the binding energy of an $S_z=2$ localized bound state calculated by iPEPS 
and the investigation of the $S_z=2$ bound state performed early on by Momoi and Totsuka. In their paper, they noticed that, thanks
to correlated hopping, two triplets can form a bound state with an energy which is minimal at ($\pi,\pi$). When measured with respect
to the energy of two isolated triplets, the energy of this state is indeed negative, as can be clearly seen in Fig.~\ref{fig:bandstructure},
which reproduces the four bands (and not just the lowest one) as calculated by Momoi and Totsuka. 
This cannot be compared to the binding of a localized bound state since at least part of the energy is gained by the coherent motion through
the lattice. Since the energy gain is due to correlated hopping, Momoi and Totsuka were led to the conclusion that the binding energy
must be entirely due to the coherent motion through the lattice, and that due to short-range repulsion, a localized bound state would
be unstable. However, even if it is localized, a bound state can gain kinetic energy by virtual local processes, and to really know whether
a localized bound state can be stable with respect to the decay into two isolated triplets, one should find a more systematic way to
estimate its energy. Once the band structure of an excitation is known, this can be achieved by constructing localized Wannier
functions from a band (or a set of bands) well separated from the other bands. Indeed, according to an old argument in one dimension due to Kohn,~\cite{Kohn59}
the localization of the Wannier function increases with the gap that separates the band (or the set of bands) from the other bands.~\cite{Blount62}
In the present case, the dispersion consists of two pairs of bands well separated from each other. So it is certainly possible to 
construct well localized Wannier functions from each pair, in particular from the lowest one. Note that there is no gap between the two
bands of the lowest-lying pair (they touch at a Dirac point), so that it is impossible to construct localized Wannier 
functions from just the lowest band. The actual calculation of well localized Wannier functions is rather involved because there is some freedom in choosing the phase
factor to optimize the localization, and it is left for future investigation. However, to get an estimate of the energy in a bound state
described by such a localized Wannier function, we do not need to actually calculate the Wannier functions. Indeed, this energy
is just the reference energy around which the dispersion is built, and in a case like here where the two lowest bands are
nearly symmetric with respect to each other, this energy is more or less equal to the average energy calculated from the two bands.
As can be seen from the middle panel of Fig.~\ref{fig:bandstructure}, this energy is clearly negative, which means that the localized
bound state described by such a Wannier function is stable with respect to the decay into two isolated triplets. Note that the scenario
proposed by Momoi and Totsuka would be realized if the average energy of the two bands was positive while the bottom 
of the lowest band is negative. This is a priori possible, but this is not the case here.

The binding energy estimated in this way is compared in the right panel of Fig.~\ref{fig:bandstructure} with the iPEPS estimate,
and the agreement is very satisfactory. Momoi and Totsuka's calculation, which is based on a perturbation calculation in $J'/J$, is
more reliable at small $J'/J$. In that limit, the binding energy estimated from Momoi and Totsuka is of the same order of magnitude
but smaller than the iPEPS estimate. This is probably due to the fact that, on a 6x6 cluster, the bound state can already gain some
delocalization energy.

\subsection{Structure of localized bound state}
\begin{figure}[htb]
\begin{center}
\includegraphics[width=0.28\columnwidth]{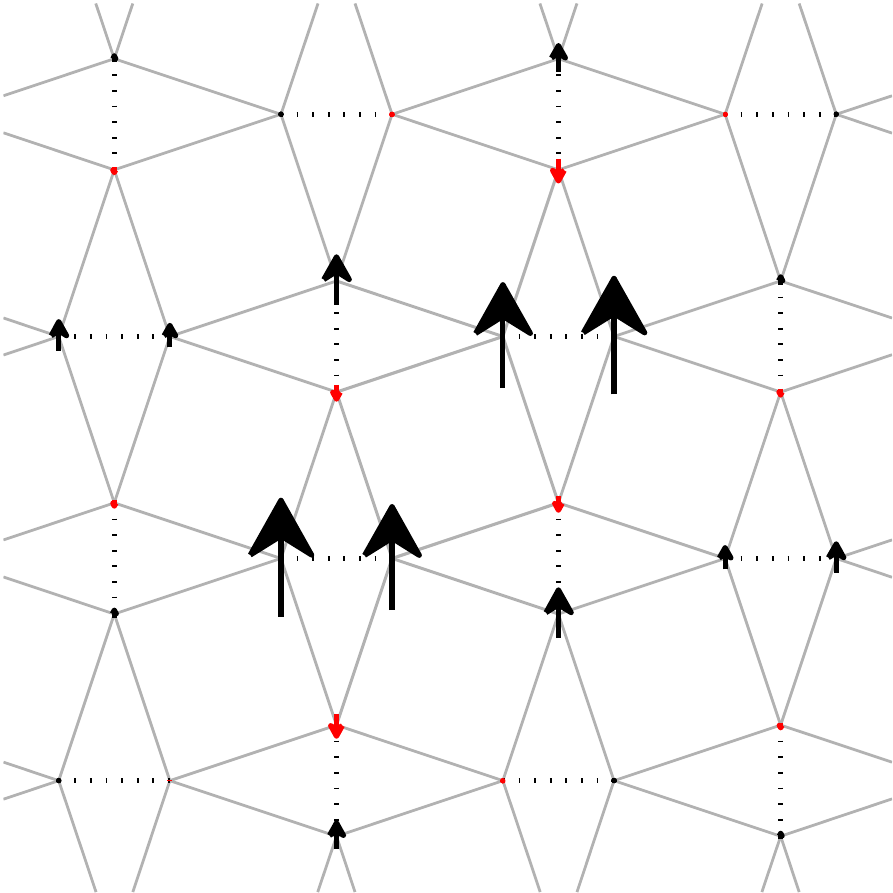} \hspace{2cm} 
\includegraphics[width=0.28\columnwidth]{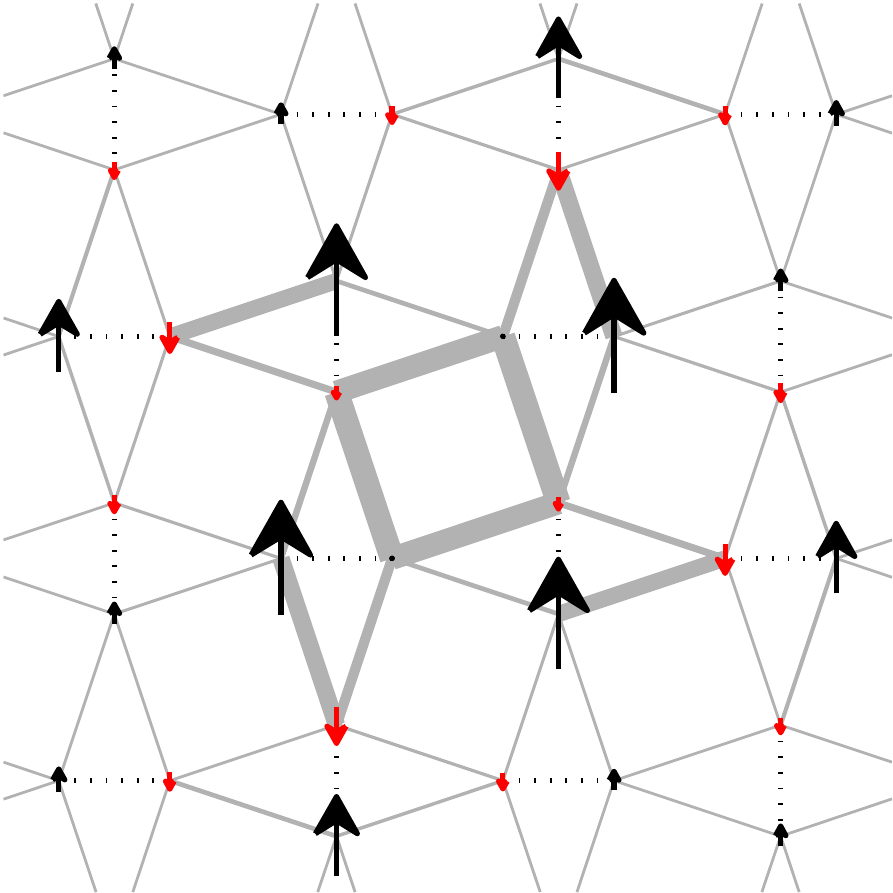}
\caption{Left panel: Spin structure of a bound state in a $4\times4$ unit cell for $J'/J=0.2$ and $D=6$ (full update). Right panel: Spin structure of a bound state in a $4\times4$ unit cell for $J'/J=0.63$ and $D=6$ (full update).}
\label{fig:comp}
\end{center}
\end{figure}

The structure of the localized bound state as revealed by iPEPS changes upon increasing the ratio $J'/J$, as shown in Fig.~\ref{fig:comp} for $D=6$. For small
$J'/J$, the bound state has only a twofold symmetry, and it essentially consists of two triplets either on the two horizontal bonds, or
on the two vertical bonds around a square plaquette, while for large $J'/J$ it acquires a four-fold symmetry and is better described as 
a singlet plaquette surrounded by four polarized spin. This evolution is again very logical in view of Momoi and Totsuka's calculation.
Up to third order in $J'/J$, the state with triplets on the two horizontal bonds is not connected to the state with triplets on the two vertical
bonds. Their calculation actually refers to only one of these states. So, as long as $J'/J$ is small, the energy gained in building
a linear combination of the two states is very small, and it is likely that other perturbations (such as the interaction with other bound
states) will rather select one of the two states, leading to the structure of the left panel of Fig.~\ref{fig:comp}. However, when $J'/J$ is large,
there will be a strong resonance between these states, leading to the four-fold symmetric pattern of the right panel of Fig.~\ref{fig:comp}.

We note that  for small $J'/J$ and a fixed bond dimension of $D=6$ the energy of the state with a twofold symmetry is only slightly lower than the state with a four-fold symmetry (of the order of $10^{-5}J$). It is conceivable that the latter state becomes energetically lower at larger $D$, even for small values of $J'/J$.

\section{Discussion of the discrepancy between  iPEPS and hierarchical mean-field theory~\cite{isaev09}}
Our iPEPS results for the plateaus below 1/4 are in disagreement with the ones obtained from a hierarchical mean-field study in Ref.~\onlinecite{isaev09}. The reasons for this discrepancy are mainly due to the cluster sizes used in Ref.~\onlinecite{isaev09} which are too small to reproduce the same crystals of bound states as we find with iPEPS. For example, to reproduce our result of a bound state in a $4\times4$ cell of dimers (1/8 plateau), clusters including 32 spins would be required, but in Ref.\onlinecite{isaev09} only clusters with 16 spins have been used for the 1/8 plateau. The same holds for the 1/6 plateau where a 12 spin cluster was used  but our proposed state for the 1/6 plateau requires clusters with 48 spins. 

\begin{figure}[htb]
\begin{center}
\includegraphics[width=0.75\columnwidth]{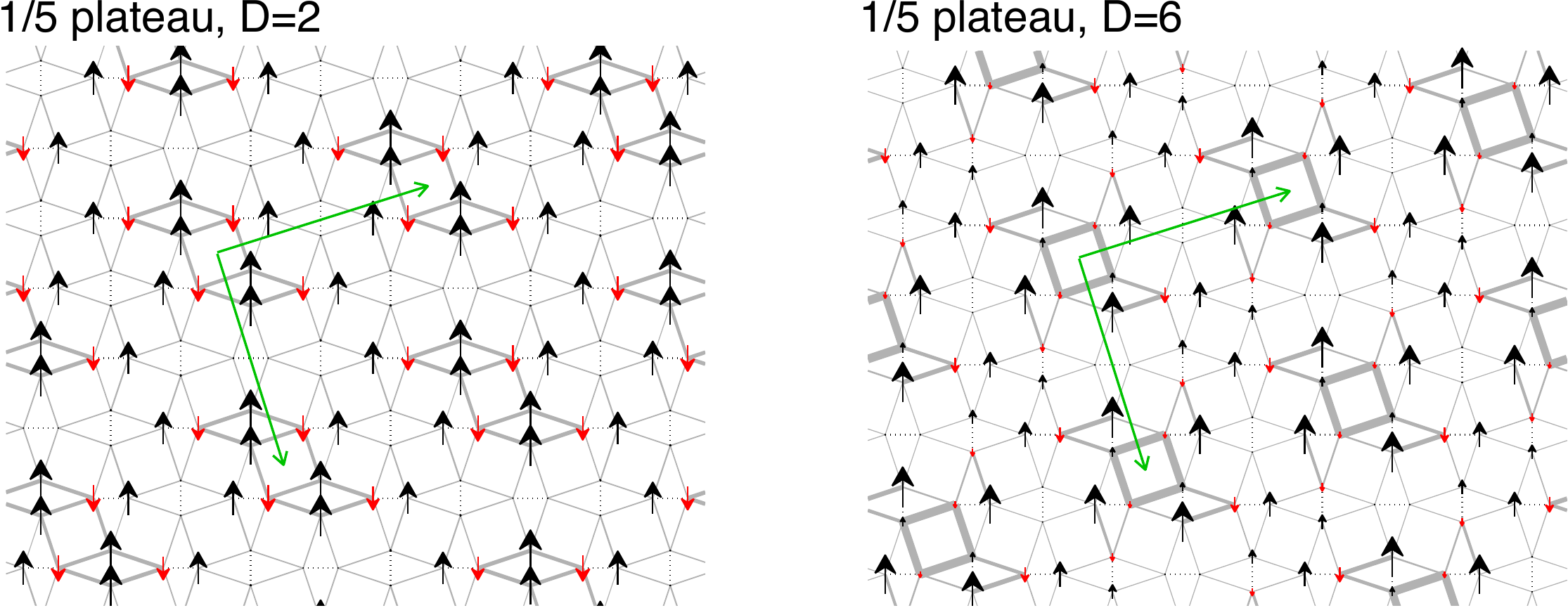} 
\caption{Left panel: Spin structure of the 1/5 plateau obtained with $D=2$. Right panel: Spin structure of the 1/5 plateau obtained with $D=6$.}
\label{fig:compar}
\end{center}
\end{figure}

However, for the 1/5 plateau the same cluster size is used  as in our calculation. Here it is interesting to see that the spin structure of the 1/5 plateau found in Ref.~\onlinecite{isaev09} is identical to the one we find for small $D=2$, see left panel of Fig.~\ref{fig:compar}. In this spin structure two triplets are touching the same plaquette. However, if we include higher-order quantum fluctuations (by going to larger $D$)  the triplets start to resonate around a plaquette, giving rise to the more symmetric  spin structure shown in the right panel in Fig.~\ref{fig:compar}. This indicates that in the hierarchical mean-field approach used in Ref.~\onlinecite{isaev09} the higher-order quantum fluctuations which create this resonance are missing. Thus, in the low-accuracy limit of iPEPS we reproduce the results from Ref.~\onlinecite{isaev09} for the 1/5 plateau, but if we increase the accuracy in iPEPS the spin structure is changed towards the more symmetric pattern of the bound states. 

\section{Triplet crystals and iPEPS}

\subsection{Sketch of the structure of some triplet patterns discussed in the text}

\begin{figure}[h]
\begin{center}
\includegraphics[width=0.8\columnwidth]{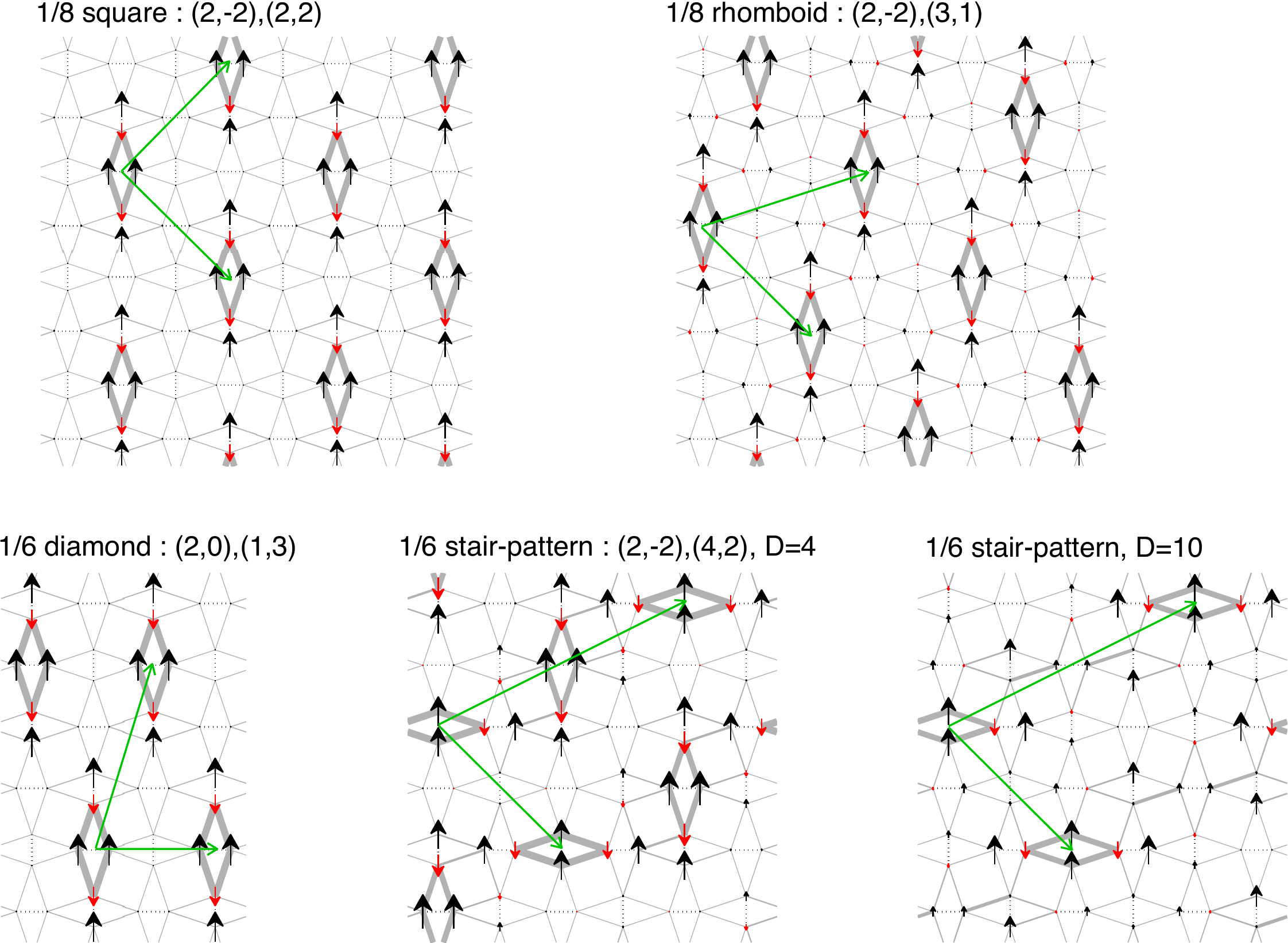}
\caption{Spin structures of the of the triplet crystals for $J'/J=0.63$, $D=10$ (simple update). The last panel shows that the 1/6 stair-pattern is not stable at large $D$.}
\label{fig:tripletcrystals}
\end{center}
\end{figure}

In Fig.~\ref{fig:tripletcrystals} we show the spin structures of the triplet crystals mentioned in the main text, obtained with  $J'/J=0.63$ and $D=10$ (simple update). Note that the 1/6 triplet crystal with stair-pattern and unit cell vectors $v_1=(2,-2)$, $v_2=(4,2)$ shown  Fig.~\ref{fig:tripletcrystals} for $D=4$ is not stable for large bond dimensions: for $D=10$ two triplets are still localized, but the other two triplets are delocalized to give rise to a different spin pattern (see last panel of Fig.~\ref{fig:tripletcrystals}).

\clearpage

\subsection{Magnetization curve resulting from triplet states}

\begin{figure}[htb]
\begin{center}
\includegraphics[width=0.6\columnwidth]{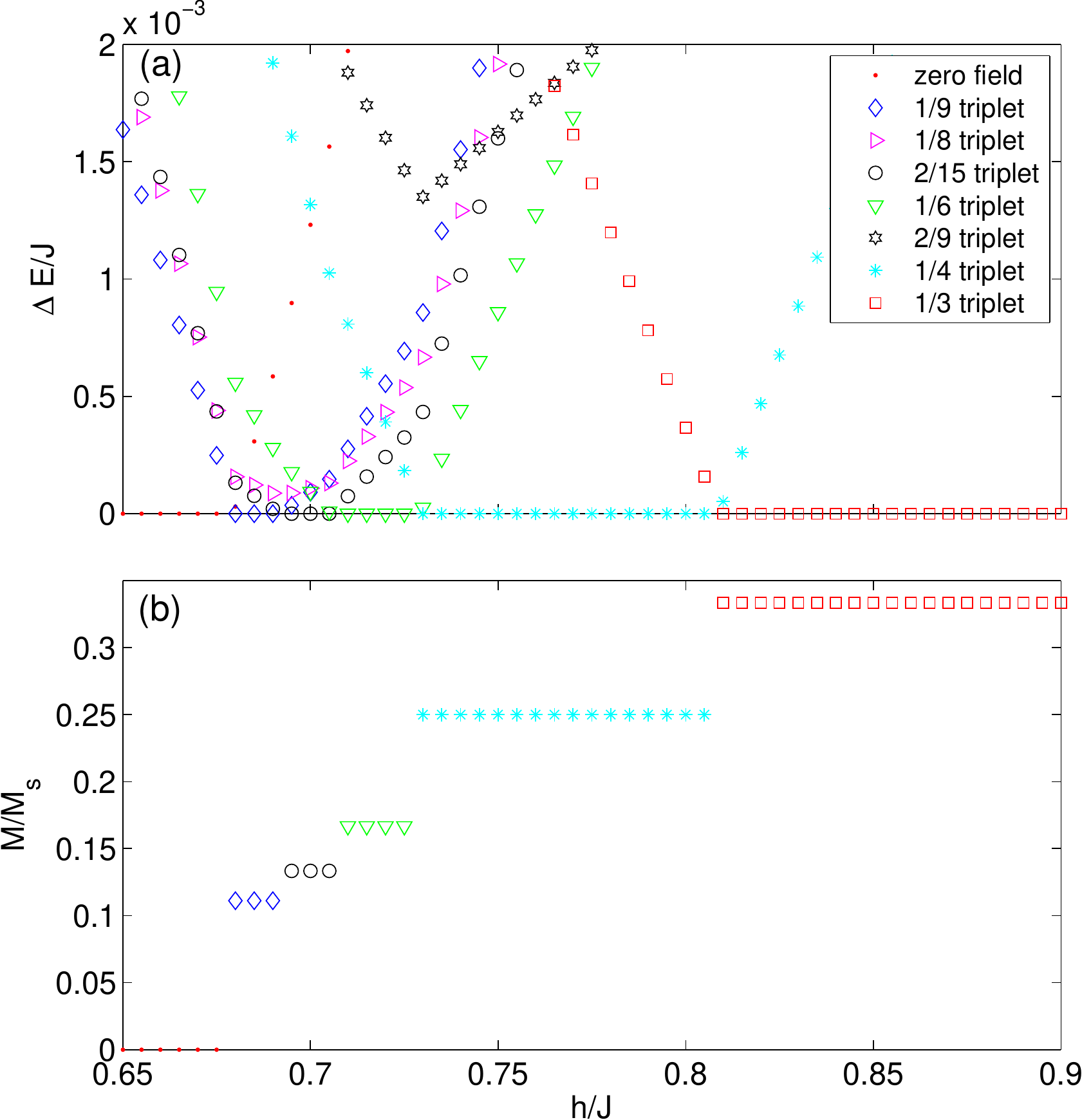}
\caption{Results for the magnetization plateaus by restricting the solution to triplet crystals for $J'/J=0.5$, $D=6$ (simple update). The triplet crystal structures are the ones from Ref.~\onlinecite{Dorier08}. 
 (a) Comparison of the variational energies of the competing triplet crystal states as a function of $h/J$ with respect to the lowest triplet crystal state. (b) The resulting magnetization curve as a function of $h/J$. }
\label{fig:res_ssmtriplets}
\end{center}
\end{figure}

In Ref.~\onlinecite{Dorier08} a sequence of magnetization plateaus (1/9, 2/15, 1/6, 2/9, 1/3) has been predicted based on the perturbative continuous unitary transformation (PCUT) approach. Here we show that we can reproduce these results (in the low-density limit) with iPEPS if we restrict our study to the same triplet crystals as considered in Ref.~\onlinecite{Dorier08}. In Fig.~\ref{fig:res_ssmtriplets}(a) we present the energy difference between the competing states, and in Fig.~\ref{fig:res_ssmtriplets}(b) the resulting magnetization curve for $J'/J=0.5$ ($D=6$). In the low-density limit of triplets ($M/M_s\le1/6$) the magnetization curve agrees with the one presented in Ref.~\onlinecite{Dorier08} (Fig.~5). In particular, the 1/8 triplet crystal is not stable with respect to the 1/9 and 2/15 triplet crystals. If we consider the energy difference plotted Fig.~\ref{fig:res_ssmtriplets}(a) we can see that the 2/9 triplet crystal becomes lower in energy than the 1/6 triplet crystal around $h/J\approx0.78$ which is also in good agreement with Ref.~\onlinecite{Dorier08}. However, we do not find that the 2/9 plateau is stable, but that the 1/4 plateau is energetically lower. Also, the 1/3 plateau starts at a smaller value of $h/J$ than in Ref.~\onlinecite{Dorier08}. This shows that in the dense limit of triplets ($M/M_s\ge1/4$), where PCUT is expected to be less accurate, the energies from Ref.~\onlinecite{Dorier08} are too high compared to the iPEPS energies. 

\bibliographystyle{apsrev4-1}
\bibliography{../../bib/refs, comments}